\documentclass[twocolumn,aps,nofootinbib]{revtex4-1}
\usepackage{amsmath}
\usepackage[english]{babel}
\usepackage{float}
\usepackage{amsfonts}
\usepackage{amsthm}
\usepackage{graphicx}
\usepackage{dcolumn}
\usepackage{stackrel,amssymb}
\usepackage{color}
\usepackage{ulem}

\usepackage{bm}

\newcommand{\ket}[1]{\vert#1\rangle}

\renewcommand{\eqref}[1]{Eq.~(\ref{#1})}
\begin{document}

%\title{Quantum synchronization-assisted machine learning of a quantum environment} 
\title{Machine learning applied to quantum synchronization-assisted probing}

\author{Gabriel Garau Estarellas, Gian Luca Giorgi, Miguel C. Soriano, and Roberta Zambrini}
\affiliation{Instituto de F\'{\i}sica Interdisciplinar y Sistemas Complejos IFISC (UIB-CSIC), UIB Campus, E-07122 Palma de Mallorca, Spain}

\begin{abstract}

A probing scheme is considered with an accessible and controllable qubit, used to probe an out-of equilibrium system consisting of a second qubit interacting with an environment. Quantum spontaneous synchronization between the probe and the system emerges in this model and, by tuning the probe frequency, can occur both in-phase and in anti-phase. We analyze the capability of machine learning in this probing scheme based on quantum synchronization. An artificial neural network  is used to infer, from a probe observable, main dissipation features, such as the environment Ohmicity index. The efficiency of the algorithm in the presence of some noise in the dataset is also considered. We show that the performance in either classification and regression is significantly improved due to the in/anti-phase synchronization transition. This opens the way to the characterization of environments with arbitrary spectral densities.\\

\end{abstract}%

\maketitle

\section{Introduction}

The development of automated procedures is one of the main goals of modern science and technology \cite{jordan}. Intelligent automation is expected to enable processes to perform such that both human and time resources are reduced and optimized. 
Among many other fields of application, machine learning (ML) algorithms are widely used in physics studies for instance to reconstruct and interpret data, for pattern
recognition, or for automated designs of new  experiments \cite{ml1,ml2}.
On top of that, quantum information technologies make available promising applications in communication and computation and are also expected to have a strong impact in our everyday life in the near future \cite{feynman,nielsen,qtech}. 
%  Scalable universal quantum computers promise to bring quantum advantage in many settings. For instance, quantum algorithms are expected to make possible speedups over the best classical counterparts for problems such as factorization \cite{shor}  and search in an unstructured database \cite{grover}. 
Quantum machine learning (QML) covers the field where  quantum computing and artificial intelligence cooperate with each other \cite{schuld,biamonte,briegel}. The interest in such discipline has raised in the last
decades and we have already experienced significant advances in both directions of influence: on the
one hand, quantum-based algorithms have been proved to speed up many ML methods;
%and also facilitate interaction between the agent and the user; 
on the other hand, ML is already being used in many cutting-edge technologies, including those based on quantum
information settings,  such as quantum design of new experiments \cite{design,design1}, quantum metrology \cite{hentschel,lumino}, Hamiltonian estimation \cite{wiebe,wang},  quantum control \cite{control}, and  the identification of phases of matter \cite{carrasquilla,venderley}.

%The emerging field of quantum machine learning (QML)  has already faced problems in quantum signal processing, quantum metrology, Hamiltonian estimation, in problems of quantum control, and in the identification of phases of matter. 

As a subfield of QML, the idea of ``quantum learning" has been introduced to design protocols that aim at 
 ``learning about'' properties of quantum systems. In some instances, one may want to learn about a quantum state or a quantum map using classical ML algorithms. Such methods have already been employed to  control and classify systems exhibiting quantum features. Examples are given by the identification of the optimal conditions to create a Bose-Einstein condensate \cite{wigley}, the detection of quantum change points \cite{sasaki}, the classification of qubits \cite{calsa}, the prediction and suppression of decoherence \cite{stenberg,mavadia,gupta}, the identification of many-body phase transitions \cite{carrasquilla,venderley}, the gate design and simulation \cite{banchi1}.

 In this paper, we propose a method to detect the fundamental properties of a large quantum system (an environment interacting with a qubit) by making use of a learning algorithm receiving the data of an observable probe, represented by another qubit. As it was reported in Ref.\cite{probing}, an out-of equilibrium qubit, interacting with an environment, can synchronize with another coupled qubit. The emergence of quantum synchronization has been recently reported in several systems  \cite{chapter} and, in particular, in spin systems
 \cite{lehur,pra,holland,ameri,solano,probing,bruno,solano2}. This phenomenon can emerge not only in the well-known case of systems exhibiting self-sustained oscillations \cite{pikov,florian,fazio}, but also during transient dynamics \cite{lehur,probing,bruno,ameri,nostri,napoli,solano}. Furthermore, the role played by dissipation and decoherence in synchronization has been explored not only in transient regimes  but also in decoherence free subspaces \cite{manzano} and in the presence of self-sustained oscillations (in opto-mechanichal systems \cite{cabot_OM}). In the case of two coupled qubits, even with different frequencies, with a significant imbalance between their losses, both synchronization and anti-synchronization can arise depending on the system parameters \cite{probing}.
 The transition between these two in-phase and  anti-phase oscillation regimes can actually be induced by tuning the frequency of the probe and can be exploited in a probing scheme: therefore,  measuring the supposed accessible and controllable probe one can infer features of the other dissipating qubit and of its environment.
 
\begin{figure}[t]
  %\centering
   \includegraphics[scale=.42]{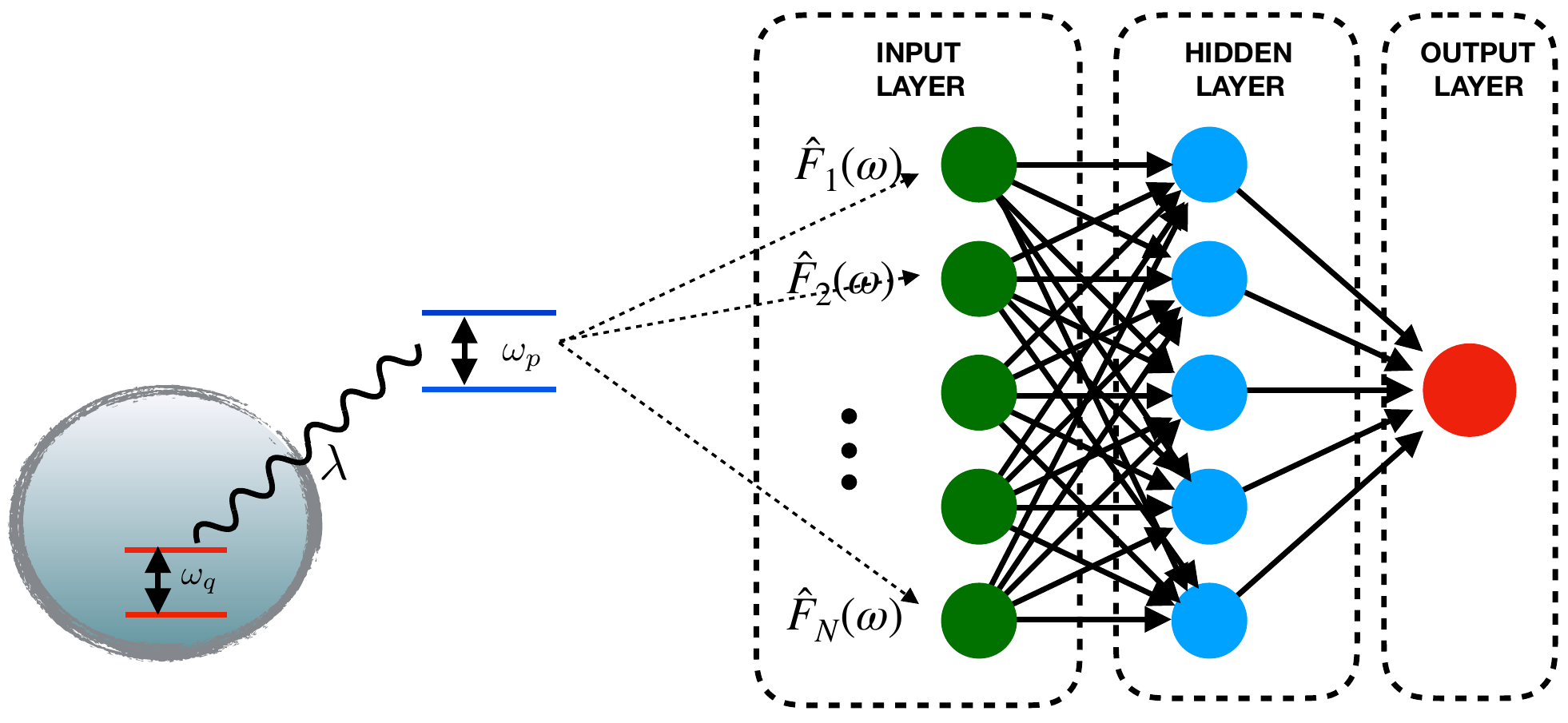}\\
 \caption{ Schematic view of the set-up. On the left part of the figure the physical system is depicted: the shadowed area represents the environment, in which the  qubit $q$ with natural frequency $\omega_q$ is immersed. The system $q$  interacts with the probe qubit $p$ (with frequency $\omega_p$) through the coupling coefficient $\lambda$. The Fourier transform $\hat{F}(\omega)$ of the measured data  $\langle\sigma_p^x(t)\rangle$ is used as  input data to train the hidden layer, where the learning takes place. Then, the output layer projects the results.
}\label{fig2}
\end{figure}

 The aim of this work is to show that the time evolution of an observable of the  probe can be used by an artificial neural network (ANN) to learn how to distinguish different forms of environments and that the transition between in-phase and anti-phase  synchronization plays any significant role.  For this purpose, as usual ANNs need to identify statistical patterns and then classify new patterns based on the previous knowledge.
 In particular, in the so-called supervised learning approach, part of the data are employed to train the ANN. The ANN uses these labelled data to learn or approximate a proper solution to the problem at hand. Then, the machine tries to assign the correct labels to the rest of the, previously unseen, data. In Figure \ref{fig2}, a pictorial view of the ML probing scheme is presented: a dissipating qubit interacts with a probe and the data of an observable of the probe are used as an input for the ANN, whose output contains information about the relevant features of the bath. 
 
 Identifying the features of any environment is essential from a technological point of view,  as the emergence of decoherence is the main detrimental effect against quantum information-processing applications.  Even when considering systems weakly interacting with the environment, the form of dissipation is determined by the spectral density: several situations are properly described by a Ohmic spectral densities  \cite{breuer,weiss,ohmic}, but many sub- and super-Ohmic deviations are also known in many solid state, biophysical, optomechanical and quantum dots systems, as recently reviewed in Ref.\cite{devega} in connection with non-Markovianity \cite{vasile}. 

 In the following sections we introduce the model and the ML approach, to provide a self-contained presentation.  In Section IV we present results of both classification and regression of the Ohmicity and damping strength characterizing different environments, discussing the role of synchronization and robustness in the presence of noise. Conclusive remarks and outlook are discussed in the last Section.

 \section{Model}
 In this section, we summarize the main results of Ref. \cite{probing} on the dynamics of the system $q$ and probe $p$ (see Figure \ref{fig2}).
A qubit $q$ dissipates in an external environment and is coupled to a second, nondissipating, qubit $p$ that is assumed to be experimentally accessible and used as an external probe. A similar probing scheme is also discussed in Ref. \cite{steve}.
The frequency of the system qubit $\omega_q$  is fixed and set as a reference while the probe frequency $\omega_p$ is tuned. The total Hamiltonian is
\begin{equation}
H=\sum_{j=p,q}\frac{\omega_j}{2}\sigma_j^z+\sum_k \Omega_k a_k^\dag a_k +\sum_k g_k (a_k^\dag+ a_k )\sigma_q^x+\lambda \sigma_q^x \sigma_p^x, \label{htot}
\end{equation}
where  $\sigma_j^i$ ($i=x,y,z$) are  Pauli matrices and $a_k$ ( $a_k^\dag$) describe the bosonic   eigenmodes of the bath  with energies $\Omega_k$ ($\hbar$ set to $1$ throughout the paper).
The dissipative process is fully determined by the spectral density of the environment $J(\omega)= \sum_k g^2_k\delta(\omega-\Omega_k)$.
Even though the environment only directly interacts  with the system qubit $q$, the  master equation for the reduced system of the two qubits must be calculated starting from the eigenstates of the full system Hamiltonian $H_S=\frac{\omega_q}{2}\sigma_q^z+\frac{\omega_p}{2}\sigma_p^z+\lambda\sigma_q^x \sigma_p^x$
whenever the system-probe coupling strength $\lambda$ is strong enough \cite{globalme,probing}.
Assuming weak dissipation, the   dynamics of the density matrix  $\rho(t)$ of the pair of qubits can be studied in the Born-Markov 
and secular approximations %\cite{bp}
with Lindblad master equation $\dot\rho(t)=-i [H_{S}+H_{LS},\rho(t) ]+{\cal D}[\rho(t)]$, where the small
Lamb shift  $H_{LS}$ commutes with $H_{S}$ %and  as small as the system-bath coupling, 
and  ${\cal D}[\rho(t)]$ is the standard dissipator   \cite{breuer}.  
The diagonalization of $H_S$ can be done employing the  Jordan-Wigner procedure, which maps spins into spinless fermions and  leads to (see Ref. \cite{probing} for the details) $H_S=E_1 (\eta_1^\dag \eta_1-1/2)+E_2 (\eta_2^\dag \eta_2-1/2)$.
%\begin{equation}
%H_S=E_1\left(\eta_1^\dag \eta_1-\frac{1}{2}\right)+E_2 \left(\eta_2^\dag \eta_2-\frac{1}{2}\right).\label{jw1}
%\end{equation}
Here, the two energies are 
 $E_1=(-\sqrt{4\lambda^2+\omega_-^2}-\sqrt{4\lambda^2+\omega_+^2})/2$ and
 $E_2=(\sqrt{4\lambda^2+\omega_-^2}-\sqrt{4\lambda^2+\omega_+^2})/2$, where $\omega_\pm=\omega_q\pm \omega_p$.

The emergence of dissipation-induced spontaneous synchronization in quantum systems has been discussed in a series of previous papers  \cite{pra,probing,chapter,nostri}. In the presence of more oscillating modes decaying at different rates, synchronization is achieved whenever there is appreciable
separation between the two largest 
decay times  characterizing the dynamics. 
Then,
slowly decaying local degrees of freedom   experience monochromatic oscillations
at the unique surviving frequency, while the relative phases among them are locked. In the model discussed here, there are two  normal modes, associated to the fermion operators $\eta_1$ and $\eta_2$ and characterized by their respective frequency oscillation $E_1$ and $E_2$, that enter the dynamics of the local observables $\sigma_p^x$ and $\sigma_q^x$. After a transient phase and taking the bath at temperature $T=0$,  we have
\begin{eqnarray}
\langle\sigma_q^x(t)\rangle &\sim & 2 \cos( \theta_++ \theta_-)e^{-\tilde\gamma_1^+t/2}
  {\rm Re}[e^{i E_1 t }\langle\tilde{\eta}_1(0)\rangle ]\nonumber\\
&+&2\sin(\theta_++\theta_-)e^{-\tilde\gamma_2^+t/2} {\rm Re}[e^{i E_2 t }\langle\tilde{\eta}_2(0)\rangle ] \label{sq1},\\
\langle\sigma_p^x(t)\rangle &\sim & 2 \sin( \theta_+ - \theta_-)e^{-\tilde\gamma_1^+t/2}
  {\rm Re}[e^{i E_1 t }\langle\tilde{\eta}_1(0)\rangle ]\nonumber\\
&+&2\cos(\theta_+-\theta_-)e^{-\tilde\gamma_2^+t/2} {\rm Re}[e^{i E_2 t }\langle\tilde{\eta}_2(0) \rangle], \label{sp1}
\end{eqnarray}
with the (fermionic) annihilation operators $\tilde{\eta}_i={\cal P}\eta_i$, where the parity operator is ${\cal P}=(1-2\eta_1^\dag\eta_1)(1-2\eta_2^\dag\eta_2)$,  and with $\theta_\pm=\arcsin(2 \lambda/\sqrt{4\lambda^2+\omega_\pm^2})/2$. The effective decay rates determining the dynamics are directly related to the spectral density and are $\tilde\gamma_1^\pm=\cos^2 (\theta+\phi)J(\pm E_1)$ and $\tilde\gamma_2^\pm=\sin^2 (\theta+\phi)J(\pm E_2)$

\begin{figure}%[h]
    \centering
           \includegraphics[scale=.5]{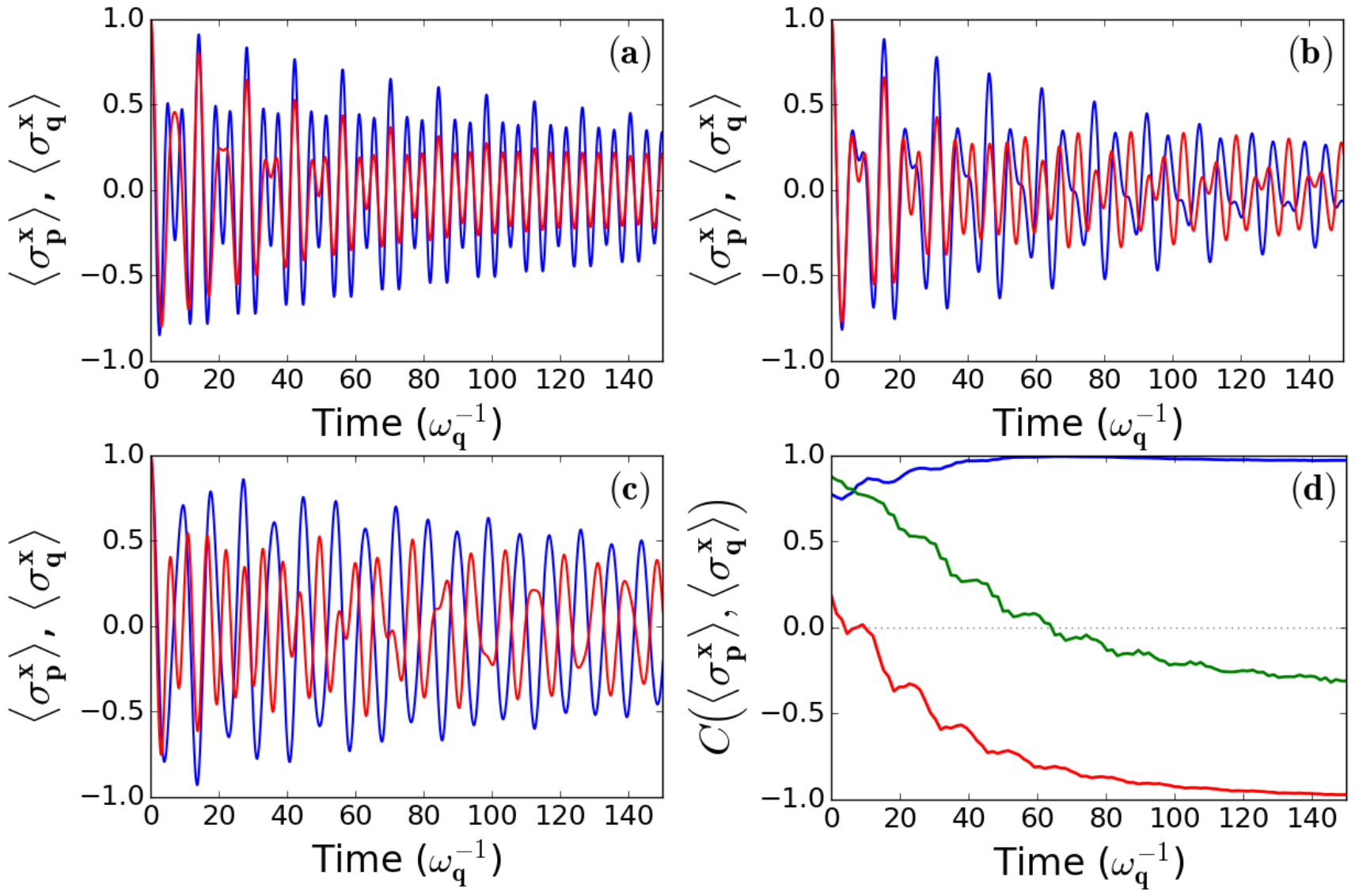}%
        \caption{Panels (a), (b), and (c):  evolution of qubit and probe exhibiting in-phase synchronization ((a) $\omega_p=1.25$ and $s=0.5$), no synchronization ((b) $\omega_p=1$ and $s=1$)
	 and anti-phase synchronization ((c) $\omega_p=0.75$ and $s=2$). Other parameters: $\gamma_0=0.01$ and $\lambda=0.2\; \omega_q$; initial state $\ket{\psi(0)}=\left(\ket{\uparrow}+\ket{\downarrow}\right)\left(\ket{\uparrow}+
	 \ket{\downarrow}\right)/2$. Blue lines represent $\langle \sigma_p^x \rangle$ and red lines   $\langle \sigma_q^x \rangle$.
	 Panel (d) shows the Pearson correlation coefficient as a function of time (setting $\Delta t=50\omega_q^{-1}$ as the width of the temporal window for its computation) for the three trajectories: the blue line corresponds to (a), the green line to (b) and the red line to (c). }
 \label{Sync}
\end{figure}

 As a result of such dynamical structure, in the long-time limit, the two qubits experience monochromatic, synchronous oscillations if either $\tilde\gamma_1 \ll \tilde\gamma_2$
or  $\tilde\gamma_1 \gg \tilde\gamma_2$. The frequency of such  synchronous oscillations is $\omega_{{\rm
sync}}\simeq E_1 $ for  $\tilde\gamma_1 \gg \tilde\gamma_2$ and $\omega_{{\rm
sync}}\simeq E_2 $ for  $\tilde\gamma_1 \ll \tilde\gamma_2$  \cite{probing}. The cases where the two decaying rates are of the same order of magnitude are characterized by the absence of synchronization, that can be quantified using the so-called Pearson correlation coefficient \cite{chapter}.
Let us now assume a power-law spectral density for the bath \cite{breuer,weiss,devega}:  
\begin{equation}\label{eqJ}
J(\omega)\sim \gamma_0\;
 \omega^s,
\end{equation}
with a high-energy cut-off, that can be consistently neglected. The condition for the absence of synchronization
$\tilde\gamma_1/ \tilde\gamma_2=1$ is satisfied along a 
line in the $\omega_p$--$s$ diagram which corresponds to 
$
\log_{ \bar E_1/ \bar E_2}\tan^2( \bar\theta_+ +  \bar\theta_-) = s$, 
where the bar indicates that all the parameters must be calculated at a given probe frequency $\omega_p=\bar{\omega}_p$. Then, determining the value of $\omega_p$ at which the transition from in-phase to anti-phase synchronization takes place amounts to estimating the value of $s$. Remarkably, such an estimation can be done by monitoring the probe alone, due to the  macroscopic jump in $\omega_{{\rm
sync}}$ that can be detected locally \cite{probing}. It is also worth noticing here that such transition does not depend on the coupling constant $\gamma_0$, which only determines the time scale of dissipation.
Examples of trajectories exhibiting in-phase, anti-phase, and absence of- synchronization are reported in Figure  \ref{Sync}, together with the respective dynamics of the  Pearson correlation coefficient.

\section{Machine learning approach to probing the environment}

The analytical solution for the probe dynamics allows one to reconstruct the shape of the spectral density  \cite{probing} but it is beneficial to apply machine-learning techniques to probe the bath properties for the following reasons: (i) the simplicity of the model permits to highlight the specific role played by the emergent phenomenon of synchronization in the automated learning procedure; (ii) since more complex environments can be either found in nature or engineered, the case discussed here can be taken as a benchmark test to explore the validity of the automated reconstruction approach. Furthermore we will also address the effect of noise in the data.

The machine-learning problem is tackled within the so-called supervised learning paradigm. The basic component is a network made by artificial neurons (ANs) distributed in one or more hidden layers that are fed with input data. Here, we used a network with a single hidden layer. Each AN is a real function parametrized by a vector of real weights $\vec{w}$ and an activation function $\varphi(\vec{x} \cdot \vec{w})$, where $\vec{x}$ is the vector of input data which the neuron receives from other neurons.  The task of the training is to optimize the weights $\vec{w}$ and, possibly, the parameters that determine the activation function $\varphi$ of each neuron, as to minimize the error in the classification of the training set. The activation function used here is the standard sigmoid function $\varphi(x)=1/(1+e^{-x})$. 

The machine-learning protocol (Figure \ref{fig2}) assists identifying the properties of the bath by measuring only the probe dynamics. The ANN is composed of three layers, namely the input layer, the hidden layer and the output layer. The input layer has $M$ nodes, which in our case correspond to the number of samples in the Fourier transform of the probe signal. The hidden layer is made up of $L$ artificial neurons, which have an activation function $\varphi(x)$. The output layer has, in our case, a single artificial neuron, which gives the final result. The connections between the layers are considered to be of feed-forward only type and the strength of these connections (weights) are optimized via back-propagation \cite{Hinton}.

 \section{Results}
 
 In the following, we consider time series of the probe interacting with an out-of equilibrium qubit in different environments, going beyond usual Ohmic ($s=1$) spectral densities. Also the probe frequency $\omega_p$ is allowed to be tuned, leading to trajectories with different dynamical evolution as shown in Figures \ref{Sync}a, \ref{Sync}b and \ref{Sync}c. 
We address the performance of the ANN in connection with the phenomenon of spontaneous synchronization, focusing on the transition between in-phase and anti-phase synchronization \cite{probing}. This is quantified by the 
Pearson coefficient \cite{chapter}, that is the correlation (normalized with variances) between the time fluctuations (with respect to the averages on a time window) of two functions. The considered functions will be the expectation values $\langle\sigma_q^x(t)\rangle$ and $ \langle\sigma_p^x(t)\rangle $.
Figure \ref{Sync}d shows the Pearson coefficient  $C(\langle\sigma_q^x\rangle \langle\sigma_p^x\rangle $) evolution for three different parameter sets that lead to either in-phase (a) or anti-phase (c) synchronization, or absence of synchronization (b).

\subsection{Classification of $s$}

We start by considering the ability to distinguish sub-Ohmic, Ohmic and super-Ohmic dissipation. We aim at identifying the proper value of $s$, which varies among three different values  ($s = 0.5,\; s = 1,\; s = 2$), for the bath power law in Equation (\ref{eqJ}) by using the ML algorithm.
To that end, we generate trajectories for different values of $\omega_p$ uniformly distributed in a certain range of frequencies nearly resonant with the one of the qubit.
The input data for the ML algorithm are the probe spectra, i.e. the modulus
%$M=101$ puntos de evolución temporal entre 0 y 100 (aunque harían falta menos, en realidad, ya que con esta discretización tenemos información hasta $\omega_{max}=1/(2\Delta t)=0.5$ y en el peor de los casos que estudiamos el segundo pico está en $\omega=0.26$) y que en cada caso hemos dado como training el mínimo número posible de puntos tal que se vieran los dos picos de la transformada.]}
of the Fourier transform $\hat{F}(\omega)$ of the time trajectories $\langle \sigma_p^x (t)\rangle$.
Proper resolution in frequency in Equation (\ref{sq1}) is achieved considering 101 points in the time interval $[0, 100 \omega_q^{-1}]$. This allows one to distinguish the sharp peaks centred around $E_1$ and $E_2$. The shape of these peaks depends on $\omega_p$ and $s$, as well as on the system-probe coupling, set to  $\lambda=0.2\; \omega_q$.
The training set consists of $N$ spectra with their respective labels (the three values of $s$) for the supervised learning procedure, while the remaining spectra (of size $0.25N$) are used for testing the performance of the algorithm. 
With this dataset, the ML algorithm must learn to classify the spectra in three different categories that correspond to the three possible values of $s$.

\begin{figure}%[h]
    \centering
           \includegraphics[scale=.5]{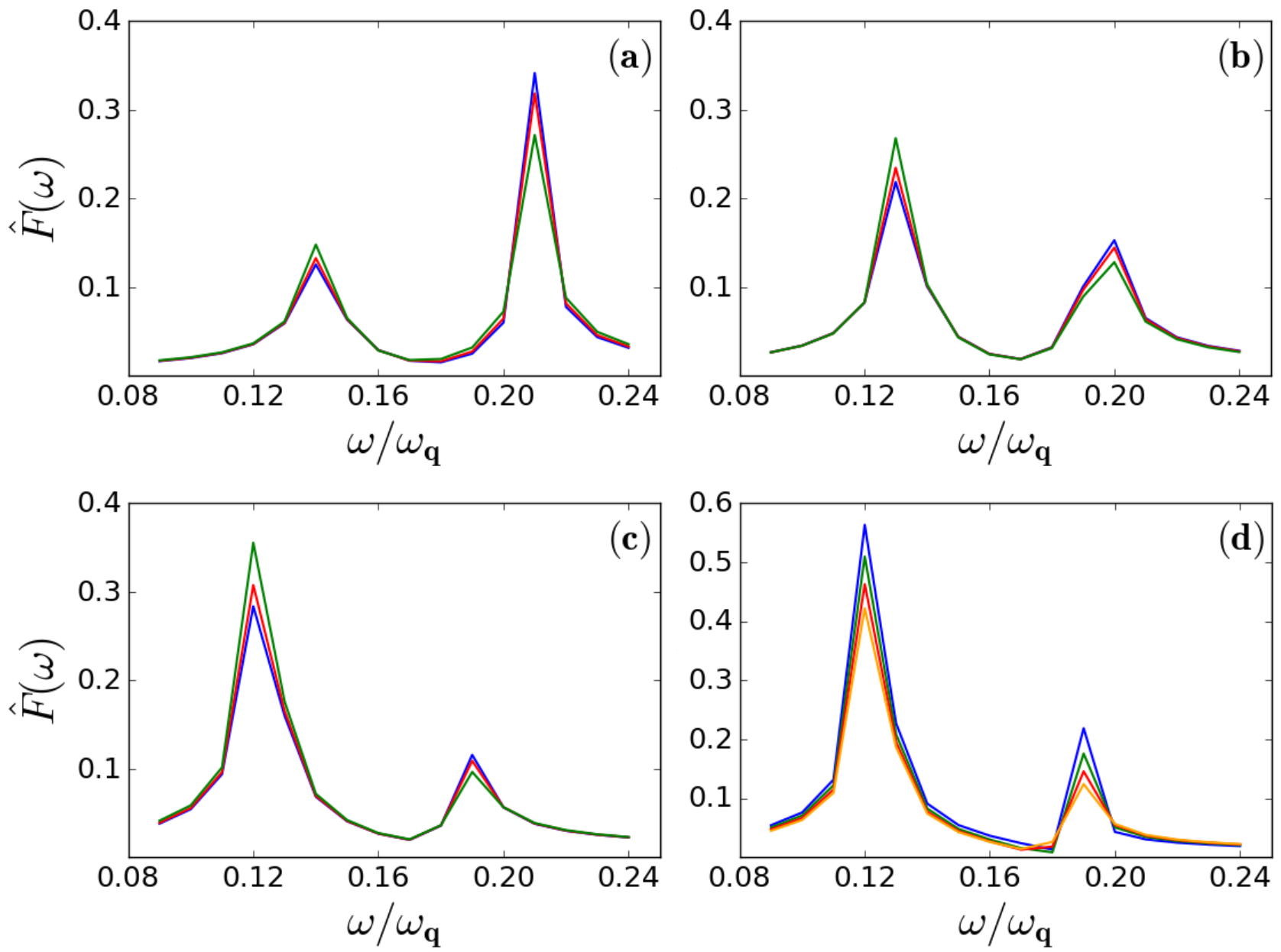}%
       \caption{Fourier transform of the probe dynamics  for (a) $\omega_p = 1.15\omega_q$ , (b)
$\omega_p =\omega_q$ and (c) $\omega_p = 0.9\omega_q$. In all cases the green line corresponds to $s=2$, the red line to $s=1$ and the blue line to $s=0.5$. 
The qubit-bath coupling coefficient is $\gamma_0 = 0.03$ in this case, in order to better visualize the differences between the trajectories for each value of $s$. Panel (d) shows the Fourier transform for a fixed value of $s=1$ and different values of the qubit-bath coupling coefficient: the blue line corresponds to $\gamma_0=0.005$, the green line to $\gamma_0=0.01$, the red line to $\gamma_0=0.015$ and the orange line to $\gamma_0=0.02$.}\label{fig1}
\end{figure}

\begin{figure}[h]
 \centering
 \includegraphics[width=8cm]{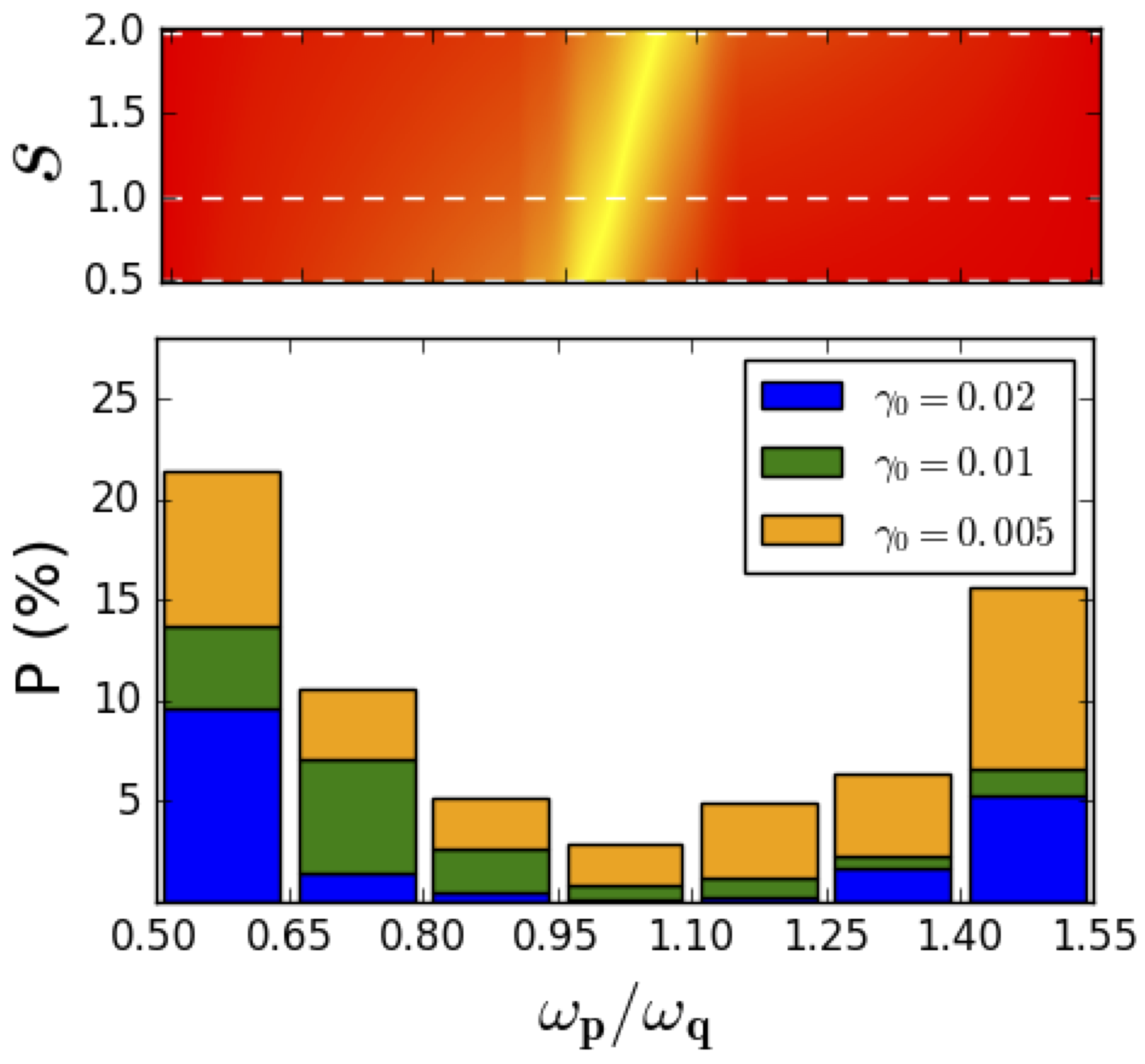}
 % 301(valores de wp)*3(valores de s)=903 sample examples, he usado 720 (240 valores de wp *3 , aprox 80%) como training y el resto  test.
 \caption{Top: absolute value of the Pearson correlation coefficient $C$, fixing $t=80\omega_q^{-1}$ and $\Delta t=20 \omega_q^{-1}$ and for Ohmicity index $s\in[0.5,2]$. The yellow band corresponds to lack of synchronization ($C \simeq 0$), while the region at the right of the band corresponds to in phase synchronization ($C>0$) and the left region corresponds to antiphase ($C<0$). Bottom: histogram of the percentage of error (that is, the percentage of incorrect labels with respect to the total number of test examples) produced by the ML algorithm in the classification between $s=0.5$, $s=1$ and $s=2$, for different intervals of $\omega_p$ and for different values of the coupling $\gamma_0$ (as in the picture). We have used $N=720$ training examples and a single hidden layer of $L=50$ neurons with the logistic activation function. The upper limits of the blue, green and orange bars correspond to the fraction of error with $\gamma_0=0.02$, $\gamma_0=0.01$ and $\gamma_0=0.005$, respectively.} \label{fig3}
\end{figure}

In this context, training the neural network means teaching it to distinguish among the family of shapes of the Fourier transform of the trajectories in order to identify the correct label in the test phase. An example of the training data is given in Figure \ref{fig1}(a-c), where  $\hat{F}(\omega)$ is plotted, for each of the $s$ to be identified and for three different values of $\omega_p$. The small changes in the probe spectrum depending on the environment ($s$) are the signature to identify the environment with the ML algorithm (we note that here only a part of the points of the spectrum are represented). 
The relative height of the peaks in Fig. \ref{fig1} encodes the dominant frequency of the system, depending on the occurrence of in-phase or anti-phase synchronization. As mentioned before, a key property of this model is indeed the existence of a phase diagram with a quite sharp transition between synchronous and anti-synchronous oscillations in the pre-asymptotic regime \cite{probing}. Considering values of $s$ between  $0.5$ and $2$, this transition takes place for $\omega_p$ approximately lying in the interval $\{0.95,\, 1.1\}$,  as shown in the top panel of Figure \ref{fig3}. Inside this region, different values of $s$ give qualitatively different dynamics (in/anti-phase synchronization), while outside it, there is a minor quantitative difference in the height and width of the peaks. 

Our results for the classification of $s$ are summarized in the lower panel of Figure \ref{fig3}. We have divided the whole training set into seven regions with different frequency ratios and implemented, for each of the seven samples, the classification protocol with the ML algorithm. Going beyond the absolute error (that can be improved with larger training sets), it is important to note the structure present in these results. We find that in the region of the spectrum where the transition in/anti-phase synchronization takes place, the classification error (that is, the rate of failure $P$ in identifying $s$) is much smaller than in the other regions, reaching a minimum. Therefore the phenomenon of synchronization in this set-up, not only allows one to probe the system, but the in/anti-phase synchronization transition facilitates the  learning process of the neural network.

This main result is explored both for very weak and less weak dissipation conditions, remaining always within the regime of validity of our model. We find that for larger $\gamma_0$, corresponding to faster  dissipation (blue color in the lower panel of Fig. \ref{fig3}), the classification error is reduced. The improvement of the ML performance for larger dissipation can be understood observing that, in the presence of bigger decay rates and considering the same time window, the effects of the environment and its characterizing features become more evident in the frequency spectra.
%more dissimilarity among the peaks arise. 

\subsection{Classification of $\gamma_0$}

We have seen that a stronger dissipation increases the ability of learning a feature of the environment such as its sub/super-Ohmic form. A different question is the ability of the ML method to discriminate among different values of the system-bath coupling $\gamma_0$. Although not graphically shown, we have observed that the ML algorithm yields low classification errors for the value of $\gamma_0$ independently of the value of $s$, with $s=0.5,1,2$. For instance a classification between  $\gamma_0=0.005,\;\gamma_0=0.01,\;\gamma_0=0.02$
is realized with errors less than $2\%$ (with $N= 720$ training examples and $L=50$ neurons). %\gl{[Supongo que hacéis referencia a los resultados que os pasé en forma de histograma pero, en lugar de eso, incluís solo una figura de mérito. Habéis puesto este número de neuronas (200) porque os dije que usé los mismos parámetros que en la clasificación de $s$ con diferentes $\gamma_0$ fijados (figura 4), sin embargo lo he comprobado y el número de neuronas que usé es 50 (ya está corregido en la caption correspondiente), perdón por el error. Por eso, si os referís a las figuras que os pasé, aquí también tendríamos que escribir 50 neuronas.]} with the logistic activation function). 
We note that the error can be optimized, with an increase in the computation time, both by using larger training sets and changing the number of hidden neurons. It is also worth mentioning that a better performance is in general obtained when considering a sample of probe frequencies uniformly (rather than randomly) distributed within the sample. Interestingly, we find that the classification is neither significantly influenced by the sub/super-Ohmic character of the bath, nor on the presence of the sharp synchronization transition between in phase and anti-phase synchronization. This can actually be expected as changing the (weak) dissipation strength leads to broader spectral line-widths but does not have strong effects on the dynamics for different $\omega_p$'s. From these results, we conclude that combining the classification of both  $\gamma_0$ and $s$, different spectra described by Eq. (\ref{eqJ}) could then be discriminated.
 
\subsection{Regression of $s$}

As a further step, instead of asking the ML algorithm to identify the correct label among a family of discrete values, we can try to estimate the value of $s$ itself through a continuous regression obtained in the limit of infinite labels. In this case, the performance of the algorithm can be tested by measuring the deviation between the true $s$ of any tested data and the output of the ANN \footnote{Regression is performed with the MLPRegressor of the module sklearn.neural network in Python.}.
\begin{figure}[b]
  \centering
   \includegraphics[scale=.5]{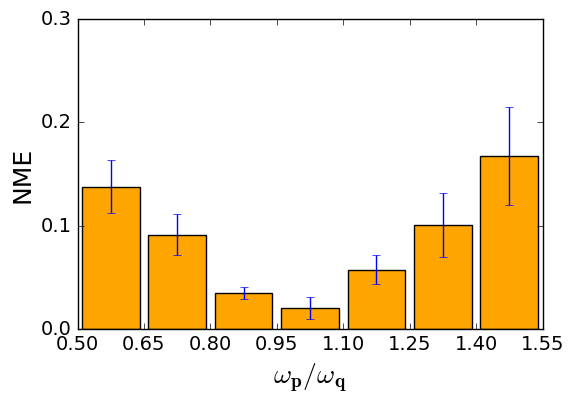}
 %31(valores de wp)*76(valores de s)=2356 sample examples, he usado 1900 (25 valores de wp *76, aprox 80%) como training y el resto test   
 \caption{Regression of $s$ for different intervals of the probe frequency. 
 The sample consists of trajectories for all values of $s \in \{0.5 + 0.02 \cdot m\}_{m=0}^{75}$ and 30 values of frequencies %$\{\omega_{min, k} + 0.005 \cdot n\}_{n=0}^{30}$ 
 in each interval. We used $L=20$ neurons and $N=1900$ training examples in each case and then calculated the error by cross-validation. The error shown is calculated as the mean value (averaged over all test examples) of the module of the differences between each of the ANN's prediction value of $s$ and the correct value of $s$ for its respective example using Eq. (\ref{error}).
 %\gl{En este caso hay menos ejemplos porque era muy pesado calcular de nuevo todas las trayectorias, como os expliqué, por eso en lugar de estar los $\omega_p$ equiespaciados en $5 \cdot 10^{-4}$ (como por ejemplo en la figura 6) lo están en intervalos de $5 \cdot 10^{-3}$, y la $s$ en intervalos de 0.02 en lugar de 0.01 (como en la figura 6).} 
 Parameters: $\gamma_0=0.01$, $\lambda=0.2$.} \label{fig4}
\end{figure}
As a figure of merit we show in Figure \ref{fig4} the performance of the regression for the identification of $s$. The normalized mean error $NME$ is calculated using the following formula:
\begin{equation}
NME=\frac{1}{T}\sum_{i=1}^T \frac{|\overline{s}_i-s_i|}{{s}_i} \label{error}
\end{equation}
being $\{s_i\}_{i=1}^T$ the correct values of the parameter that is being predicted (in this case, $s$) and $\{\overline{s}_i\}_{i=1}^T$ the respective prediction values of the ANN, for each of the $T$ test examples in the considered interval. We find that, similarly to the classification results shown in Fig. \ref{fig3}, the error in the regression is significantly reduced in the region of the in/anti-phase synchronization transition. The error bars for the NME in each frequency interval in Fig. \ref{fig3} show that the results are robust when considering the variance in the performance by cross-validation, i.e. different combinations of training and testing datasets. The variance in the regression error is also larger far from the in/anti-phase synchronization transition. 
It is worth mentioning that the results presented in Fig. \ref{fig3} do in general worsen if the time trajectories, instead of the Fourier spectra, are used as input data to the ML algorithm. We also find that the regression results worsen if only reduced information of the spectrum is used, as for instance the values for the peak positions and heights. \\

After training the ML algorithm, the weights connecting the input and the hidden layers have been optimized in order to minimize the regression error. Interestingly, we have observed that the optimized weights capture the structure of the frequency spectra that feed the ANN. More specifically, the weights associated to the frequencies around the two main peaks of the spectra have a larger absolute value. This indicates that the ML algorithm has indeed identified the most relevant features of the input data.\\

\begin{figure}[H]
 % 501(valores de wp)*150(valores de s)=75150 sample examples, he usado 15000 (100 valores de wp, aprox 20%. He usado 12000 training examples, ya que he reducido la sample total a una quinta parte (75000/5=15000 ejemplos) y he escogido el 80% como training y el resto como test
   \includegraphics[scale=.5]{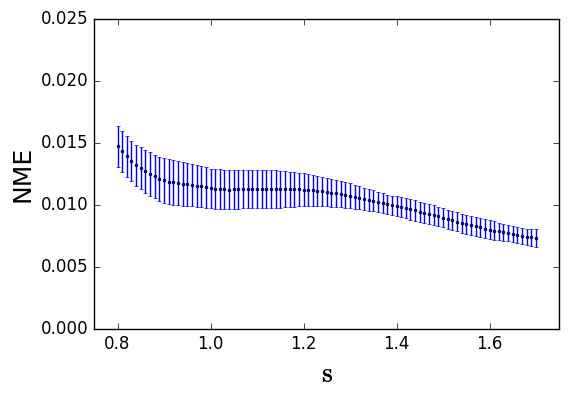}
 \caption{NME in the regression of $s$ as a function of $s$. The sample consists of $\omega_p \in \{0.9+0.0005\cdot i\}_{i=0}^{500}$ and $s$ values spaced by $0.02$.  We have used $N=12000$ training examples and $L=50$ neurons for a fixed $\gamma_0=0.01$. The error bars represent the error obtained by cross-validation.}   \label{fig4b}
\end{figure}

Finally, one could wonder if it is easier to infer the features of the environment when it is Ohmic. In order to address this issue, we look at the results of the regression but now as a function of the Ohmicity parameter $s$. Figure \ref{fig4b} shows the regression $NME$ for $s$ taking the probe frequency in the interval around the synchronization phase transition ($\omega_p \in \{0.9\; \omega_q,1.15\; \omega_q\}$). Interestingly, we find that there is no significant difference in the performance when considering environments other than the Ohmic one. Here, we have considered a large number of examples for the training set such that we obtain estimated values of $s$ with a precision of around $1\%$.  The error bars  obtained by cross-validation are also displayed.

\subsection{Machine learning in the presence of noise}

\begin{figure}[h]
\centering
\includegraphics[scale=.5]{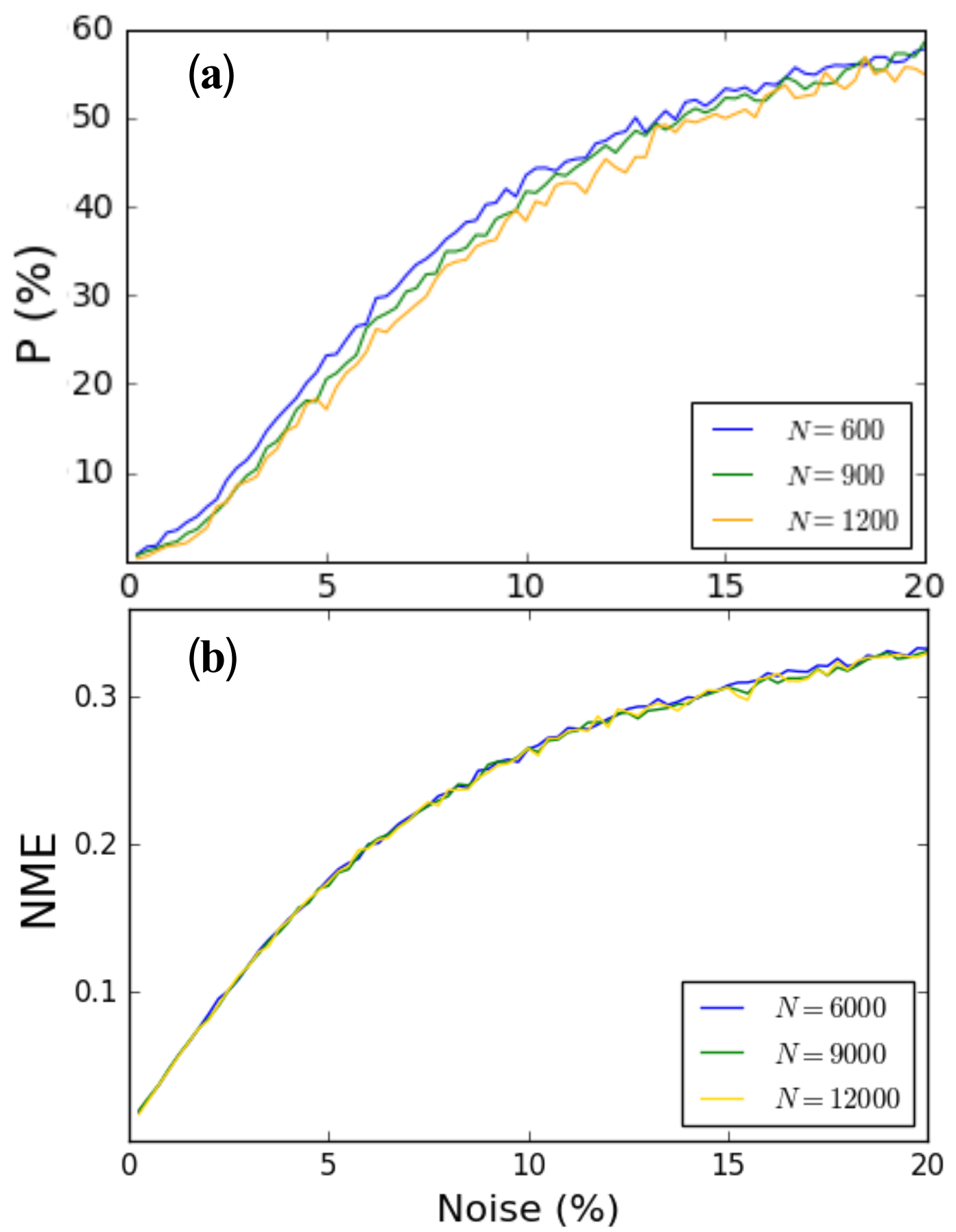}\\
\caption{Error produced in the presence of noise as a function of the percentage of noise added to the trajectories. Panel a): Probability of failure in classification. Panel b): NME in regression. We have calculated the trajectories for all values of $\omega_p \in \{0.9+0.0005 \cdot i\}_{i=0}^{500}$, and used for ML in all cases $L=50$ neurons and the number of training examples $N$ specified in the legend. The percentage of noise represents the relation between its amplitude (understood as the standard deviation) and the total amplitude of the function $\langle\sigma_p^x(t)\rangle$. Other parameters: $\gamma_0=0.01$ and $\lambda=0.2$.}
\label{Noise}
\end{figure}

An important question when considering possible experimental implementations of the current scheme is whether the ANN can identify the features of the out-of equilibrium spin system when the time trajectories are noisy. In this section, we report on the effect of an increasing noise strength both in the classification and regression of $s$. Considering the results shown in the previous sections, we focus on the probe frequency leading to a transition from in phase to anti-phase synchronization. The results shown in Figure \ref{Noise}, obtained for $\omega_p \in \{0.9\; \omega_q,1.15\; \omega_q\}$, illustrate that the performance monotonically degrades with an increasing noise strength, as one may have expected. The rate of error in classification (Figure \ref{Noise}a) reaches $10\%$ for a Gaussian noise of amplitude of $3\%$ of the trajectory. A qualitative similar trend is found in the regression error when increasing the strength of the noise, Figure \ref{Noise}b.

%
%\begin{figure}[t]
%\centering
%\includegraphics[height=50mm]{Error_vs_Noise_L_50.png}\\
%\includegraphics[height=50mm]{Error_vs_Noise_Reg_NME.png}
%\caption{Error produced in the presence of noise as a function of the percentage of noise added to the trajectories. Top: Probability of failure in classification. Bottom: NME in regression. We have calculated the trajectories for all values of $\omega_p \in \{0.9+0.0005 \cdot i\}_{i=0}^{500}$, and used for ML in all cases $L=50$ neurons and the number of training examples $N$ specified in the legend. The percentage of noise represents the relation between its amplitude (understood as the standard deviation) and the total amplitude of the function $\langle\sigma_p^x(t)\rangle$. Other parameters: $\gamma_0=0.01$ and $\lambda=0.2$.}
%\label{Noise}
%\end{figure}
%

The saturation in the performance for large noise strengths approaches the statistical limit, dictated by either the number of labels in the classification or the range of values of $s$ ($[0.5,2]$ in the regression. Furthermore, our results show that the performance worsens with noise independently of the number of examples as far as regression is considered, while a slight performance improvement can be obtained in classification when increasing the size of the training set.  

\section{Summary and outlook}

 The presence of dissipative two-level systems is ubiquitous in many experimental set-ups and the precise knowledge of the unavoidable dissipative processes is of paramount importance \cite{platforms}. Such systems are also  suitable to be used in the framework of quantum simulations in a variety of platforms  and especially using ultracold atoms in optical lattices \cite{lewenstein}. Their control represent a fundamental tool towards the development of quantum technologies. In this context, it is relevant to establish probing schemes that allow for a proper characterization of the dissipation features of qubits.

 Building on the proposal for quantum probing of Ref. \cite{probing},
we have studied here its performance using a ML learning estimation of the environmental properties.  The learning process is implemented using a neural network which is supplied by the Fourier series of the probe dynamics taken for different values of the frequency of the probe itself and of the bath spectral density. In the language commonly employed in robotics, the ANN plus the probe qubits can be seen as our intelligent ``agent" that tries to learn from and adapt to the ``environment" (here the word environment should not be confused with the quantum bath causing dissipation on the system qubit).

The main result of our work is that such a learning procedure is strongly enhanced by quantum synchronization, namely by the presence of a sharp transition between in-phase and anti-phase oscillations in the pre-asymptotic regime. In fact,
both in  classification  and in  regression of features of the bath, such as its Ohmicity parameter, 
the information about the synchronization phase strongly reduces the amount of estimation errors generated by the algorithm, making it very efficient. We have also studied how the presence of noise (unavoidable in experimental set-ups) affects the learning precision and found that the algorithm is robust even in the presence of
 (moderate) noise, both in classification and regression.

Once the advantage played by the synchronization  has been proved in our benchmark case, Eq. (\ref{eqJ}), more complicated scenarios can be also tested with arbitrary spectral densities of the dissipating qubit and also going beyond the weak-coupling regime where deviations from Ohmicity become especially important. This is relevant in experimental platforms where  there is no prior information about the spectral density \cite{probing,devega,eisert}.

\begin{acknowledgments}

We acknowledge financial support by MINEICO/AEI/FEDER through the project EPheQuCS FIS2016-78010-P. GG acknowledges funding from the IFISC SURF program, GLG from the Conselleria d'Innovaci\'{o}, Recerca i Turisme del Govern de les Illes Balears, and MCS from MINECO through a ``Ramon y Cajal'' Fellowship (RYC-2015-18140).

\end{acknowledgments}

\end{document}